\documentclass[%
 reprint,
superscriptaddress,
 amsmath,amssymb,
 aps,
pra,
longbibliography
]{revtex4-2}

\usepackage{graphicx}
\usepackage{dcolumn}
\usepackage{bm}
\usepackage{pdfpages}
\usepackage{etoolbox} 

\makeatletter
\patchcmd{\@outputpage@head}{\@ifx{\LS@rot\@undefined}{}{\LS@rot}}{}{}{}
\makeatother

\begin{document}

\preprint{APS/123-QED}

\title{Environment-assisted quantum transport and mobility edges}

\author{Donny \surname{Dwiputra}}
 \email{donny.dwiputra@s.itb.ac.id}
 \affiliation{Theoretical Physics Laboratory, Faculty of Mathematics and Natural Sciences, Institut Teknologi Bandung, Jl. Ganesha 10, Bandung 40132, Indonesia}
\author{Freddy P. \surname{Zen}}
\email{fpzen@fi.itb.ac.id}
 \affiliation{Theoretical Physics Laboratory, Faculty of Mathematics and Natural Sciences, Institut Teknologi Bandung, Jl. Ganesha 10, Bandung 40132, Indonesia}
 \affiliation{Indonesian Center for Theoretical and Mathematical Physics (ICTMP), Bandung 40132, Indonesia}

\date{\today}

\begin{abstract}
Environment-assisted quantum transport (ENAQT) is a unique situation where environmental noise can, counterintuitively, enhance the transport of an open quantum system. In this paper, we investigate how the presence of a one-dimensional single-particle mobility edge can generate strong ENAQT. For this purpose, we study the energy current of a generalized Aubry-Andr\'e-Harper (AAH) tight binding model coupled at its edges to spin baths of differing temperature and dephasing noise along the system. We find that the ENAQT increases by orders of magnitude and depends on the number of localized eigenstates and disorder strength nonmonotonically. We show that this enhancement is the result of the cooperation between population uniformization and localization. 
\end{abstract}

\maketitle


\section{INTRODUCTION}
Understanding and controlling the quantum transport of charge and energy are at the forefront of research in the field of quantum chemistry and thermodynamics. One particular phenomenon which has attracted considerable attention, both theoretically \cite{mohseni2008environment,plenio2008dephasing,rebentrost2009environment,rebentrost2009role,chin2012coherence,kassal2012environment,li2015momentum,caruso2009highly,dutta2017environment,zerah2018universal,dwiputra2020driving,dwiputra2019driving,zerah2020effects} and experimentally \cite{sowa2017environment,biggerstaff2016enhancing,trautmann2018trapped,schempp2015correlated,maier2019environment}, are the environment-assisted quantum transport (ENAQT) where moderate dephasing noise can enhance transport efficiency. Fueled by the quest for sustainable energy, several studies have exploited the potential benefit from the interplay between coherence and dissipation in quantum devices such as to enhance the power of solar cells or quantum heat engines \cite{dorfman2013photosynthetic,scully2011quantum,damanet2019controlling,streltsov2017colloquium}. These operation devices may rely on mobility edge, which is the energy threshold separating the localized and extended wavefunctions in a lattice. However, the effect of mobility edge to ENAQT is far from clear.

ENAQT may be explained by different mechanisms: initially its origin is understood from the destruction of Anderson localization \cite{anderson1958absence}---which hinders transport---by dephasing in disordered systems \cite{mohseni2008environment,rebentrost2009environment,rebentrost2009role,plenio2008dephasing,rebentrost2009environment,chin2013role} such as in photosynthetic complexes \cite{lambert2013quantum,rey2013exploiting,ai2013clustered,tao2020coherent,kim2021quantum,ishizaki2009adequacy,ishizaki2009unified,panitchayangkoon2010long,chin2013role,berman2015superradiance,duan2017nature,wang2018efficient,jang2018delocalized,yang2020steady,cao2020quantum,harush2021photosynthetic,higgins2021photosynthesis}, in which the role of quantum effects is currently under debate. In this context, ENAQT attempts to elucidate the subject.  Yet, it persists even in ordered systems, where localization does not play a role, and in fact it is impossible \textit{only} for transport in ordered chains \cite{kassal2012environment}. Thus, it relies on beneficial competition between coherent and incoherent dynamics which results in higher population entering the sink. In some specific cases, momentum rejuvenation \cite{li2015momentum}, line-broadening \cite{caruso2009highly}, and superradiance \cite{berman2015superradiance} may additionally explain the high efficiency. Moreover, exposing the system to a periodic driving may increase the efficiency even further \cite{dwiputra2020driving,dwiputra2019driving}. It is only revealed recently that the uniformity of steady state population plays more universal role in ENAQT \cite{zerah2018universal}. Dephasing noise leads to decoherence in site basis, which in turn reduces fluctuation and thereby uniformize the population and brings more particle to the sink. For end-to-end transport, the population spread is already maximized; dephasing will only reduce the current unless static disorder or localized eigenstate exists.

The interesting consequence of Anderson localization is the quantum phase transition between extended (metallic) and localized (insulating) states as a function of disorder strength. In one and two dimensions, all eigenstates are localized for a random (uncorrelated) disorder of any strength \textit{independent} of energy while in three dimensions the localization transition occur in a critical disorder strength forming a sharp mobility edge. Yet if we replace the random disorder with a quasiperiodic potential such as the paradigmatic Aubry-Andr\'e-Harper (AAH) model \cite{aubry1980analyticity,harper1955single}, which has been realized in recent ultracold atom experiments with interacting and noninteracting particles \cite{schreiber2015observation,luschen2017observation,luschen2018single}, the localization transition occurs even in one dimension. Interestingly, by introducing nonlocal hopping terms or deforming the on-site potential, the model can manifest an exact mobility edge \cite{biddle2011localization,biddle2010predicted,biddle2009localization,ganeshan2015nearest}.

In this paper, we show that a model with single particle mobility edge (SPME) can manifest a strong ENAQT, measured by comparing the energy current to the one without dephasing. Note that we do not intend to extend the understanding of the biological light harvesting \textit{in vivo}. Instead, we aim to demonstrate the transport enhancement brought by ME, which may be implemented in state-of-the-art experiments. We investigate the role of population uniformization mechanism together with the ME localization parameters to the ENAQT. We will show that the maximum ENAQT varies nonmonotonically with the ME deformation parameter and disorder strength. The enhancement also can be tuned further by varying the phase and the dephasing temperature. To isolate the ENAQT from the enhancement due to interference effects at the exit site, we consider an end-to-end transport.

\section{MODEL}
We study a one-dimensional generalized AAH (GAAH) model \cite{ganeshan2015nearest} of $N$ bosonic modes with nearest-neighbor hopping described by the Hamiltonian 
\begin{equation}\label{H_S}
	H_S = -t\sum_{n=1}^{N-1} (a_n^\dag a_{n+1} + \mathrm{H.c.}) + \sum_{n=1}^N V_n a_n^\dag a_n,
\end{equation}
where $a_n (a_n^\dag)$ is the usual bosonic annihilation (creation) operator at site $n$, $t$ is the tunneling rate, which sets our energy unit. We consider zero and single-excitation manifolds given by the states $|0\rangle$ and $|n\rangle=a_n^\dag|0\rangle$, so that the local modes $a_n = |0\rangle\langle n|$. The on-site potential is
\begin{equation}
	V_n = \Lambda + 2\lambda \frac{1-\cos(2\pi\beta n+\phi)}{1+\alpha\cos(2\pi\beta n+\phi)}.
\end{equation}
Here $\lambda$ sets the average disorder strength and the offset is chosen to be $\Lambda=2$ to avoid negative eigenvalues. We consider the mobility edge parameter $\alpha\in(-1,1)$, which can be tuned to represent several limiting cases: $\alpha=-1$ corresponds to an ordered chain, $\alpha=0$ is the original (scaled) AAH model, and $\alpha=1$ contains singular potential. We set $\beta=(\sqrt 5 -1)/2$ for a quasiperiodic modulation. This model has a SPME at $E_\mathrm{ME}=2\mathrm{sgn}(\lambda)(|t|-|\lambda|)/\alpha + \Lambda$. In the following, we use units for which $\hbar=k_\mathrm{B}=1$, and $t=1$ as the unit of energy.

The chain is coupled to two different environments: pair of thermal baths of noninteracting spins $\{\sigma_{\mu,l}\}$ at different temperatures to its edges $l=\{1,N\}$ ($B_1$), which induces current, and \textit{local} dephasing noise due to spin bath coupled to all sites $n=\{1,\dots,N\}$ ($B_2$). We explicitly considers the dynamics of the current heat baths and the interactions using the global master equation \cite{gonzalez2017testing,hofer2017markovian} to guarantee a thermodynamically consistent model \cite{gelbwaser2017thermodynamic}---in contrast to models where particle is initiated in one site and transferred to an irreversible sink. The method is generally valid for environmental rates much smaller than $t$ and sysems without quasiresonant levels. The total Hamiltonian is $H=H_S+\sum_{i=1}^2 (H_{B_i}+H_{SB_i})$, where $H_{B_1}=\sum_{l,\mu} \varepsilon_{\mu,l} \sigma^z_{\mu,l}$, $H_{B_2}=\sum_{n,\mu} \varepsilon_{\mu} \sigma^z_{\mu,n}$, and the interactions
\begin{eqnarray}\label{Hint}
	H_{SB_1} &=& \sum_{l,\mu} g_\mu^{(1)} (a_l+a_l^\dag)(\sigma^+_{\mu,l}+\sigma^-_{\mu,l}), \\
	H_{SB_2} &=& \sum_{n,\mu} g_\mu^{(2)} a_n^\dag a_n (\sigma^+_{\mu,n}+\sigma^-_{\mu,n}).
\end{eqnarray}
We choose spin baths because current rectification, which we expect to increase transport directivity, is possible if the system and bath have different statistics \cite{balachandran2019energy,saha2019particle}. Nevertheless, bosonic baths will also produce similar but smaller energy current.

The derivation of the Lindblad master equation \cite{gorini1976completely,lindblad1976generators} is most readily done using the eigenoperators method \cite{breuer2002theory,[{For derivation of master equation in similar systems, see e.g. in the Appendix of }] guimaraes2016nonequilibrium} (see Appendix \ref{app:1} for details). In the eigenbasis $|\eta_k\rangle=\sum_n S_{nk}|n\rangle$, the system Hamiltonian becomes $H_S=\sum_k \epsilon_k \eta^\dag_k\eta_k$ and the local mode is $a_n=\sum_k S_{nk}\eta_k$, where $\epsilon_k$ is the $k$-th energy (arranged in increasing order) and $\eta_k$, $\eta_k^\dag$ being its eigenoperators. In terms of the eigenoperators of $H_S$, we can write for each interaction $H_{SB_{(1,2)}}=\sum_{\omega,k} A_{i,k}(\omega) \otimes B_{i}$ where $i=\{l,n\}$ is the site index for $SB_1$ and $SB_2$ baths, respectively. Here the decomposed operator $A_{i,k}(\omega)\equiv\sum_{\epsilon'-\epsilon=\omega}\Pi(\epsilon)A_{i,k}\Pi(\epsilon')$ acts on the system, where $\Pi(\epsilon)$ is the projector onto eigenspace belonging to the eigenvalue $\epsilon$ of $H_S$, and  $B_{i}$ acts on the bath. The operator $A_{i,k}(\omega)$ is chosen to satisfy $	[H,A_{i,k}(\omega)]=-\omega A_{i,k}(\omega)$. It follows that, for the respective baths,
\begin{eqnarray}
	A_{l,k}(\omega)&=& S_{lk}\eta_k\; \delta_{\omega,+\epsilon_k}+S_{kl}^*\eta_k^\dag\; \delta_{\omega,-\epsilon_k}, \\
	A_{n,k}(\omega)&=& S_{nk} S_{nk'} \eta_k^\dag\eta_{k'}\; \delta_{\omega,\epsilon_{k'}-\epsilon_k}.
\end{eqnarray}
while $B_i = \sum_\mu g_\mu^{(1,2)} (\sigma_{\mu,i}^+ + \sigma_{\mu,i}^-)$ for both $i=\{l,n\}$. Intuitively speaking, these operators induce transitions in the system with the allowed energies $\omega$. To proceed, we use standard method to calculate the spectral densities $J_i(\omega)$ for $B_i$. The resulting master equation is 
\begin{eqnarray}\label{master}
	\nonumber &\dot{\rho}& = - i[H_S,\rho]
	+\sum_{l,k}|S_{lk}|^2 J_1(|\epsilon_k|)\Big\{(1-n_l(\epsilon_k))\mathcal D[\eta_k]\rho \\ &+& n_l(\epsilon_k)\mathcal D[\eta_k^\dag]\,\rho\Big\} 
	+ \sum_{n,k,k'} \tilde\gamma_{kk'} |S_{nk}|^2|S_{nk'}|^2 D[\eta_k^\dag\eta_{k'}]\,\rho,
\end{eqnarray}
where $\mathcal D[A]\rho = A\rho A^\dag - 1/2\{A^\dag A,\rho\}$ is the dissipator, the rate $\tilde\gamma_{kk'}=J_2(\omega)\big(1+n_\gamma(\omega)\big)+J_2(-\omega)n_\gamma(-\omega)$ is for the dephasing bath with $\omega=\epsilon_{k'}-\epsilon_k$, and $n_i(\omega)=(e^{\omega/T_i}+1)^{-1}$ is the spin occupation of the baths where $T_h,T_c$ are the temperatures for both ends and $T_\gamma$ is the dephasing temperature. These superoperators describe a nonequilibrium condition in which energy is constantly pumped and absorbed at both ends while each site is being dephased. Since we have no information about the system-bath coupling, we assume for simplicity that $J_1(\omega)=1$; yet ohmic spectral density, $J_1(\omega)\propto\omega$, will produce similar results. We choose $J_2(\omega)=\gamma$ for $\omega>0$ and $J_2(\omega)=0$ elsewhere to describe the dephasing rate.

The nonequilibrium steady-state (NESS) can be calculated exactly once the Hamiltonian is diagonalized. It is found by solving the system of equations:
\begin{equation}\label{NESS}
	A_k\langle \eta_k^\dag\eta_k \rangle - \sum_{i=1}^N C_{ki}\langle \eta_i^\dag\eta_i\rangle= B_k,
\end{equation}
with the coefficients
\begin{eqnarray}\label{coefs}
	\nonumber \centering A_k &=& \sum_l |S_{lk}|^2 (1-2n_l(\epsilon_k)) + \sum_{i} C_{ik}, \\
	B_k &=& \sum_l |S_{lk}|^2 n_l(\epsilon_k), \; C_{ik}=\tilde\gamma_{ik} \sum_n |S_{ni}|^2|S_{nk}|^2,
\end{eqnarray}	
while the off-diagonal components are zero. Note that this state needs to be normalized. The peculiar factor $-2 n_l$ appears because the system and the bath have different statistics; it vanishes if we use bosonic baths instead. Dephasing contributes to the coefficient $C_{ik}$, which measures the coupling between two eigenmodes. We consider the energy current \cite{wu2008energy},
\begin{eqnarray}\label{J} 
	\nonumber \mathcal J &=& -\mathrm Tr[H\mathcal D_N\, \rho] \\
	& = & \sum_k \epsilon_k |S_{Nk}|^2 \left(\left(1-2n_N (\epsilon_k)\right) \langle \eta_k^\dag\eta_k \rangle - n_N(\epsilon_k)\right),\quad\,
\end{eqnarray}
where $\mathcal D_N \cdot$ is the sum of $SB_1$ superoperators acting on the exit site $N$. 

\section{RESULTS}
To capture both ENAQT and mobility edge features, we choose a relatively short chain of $N=22$ so that dephasing does not completely suppress the current, but still large enough to pronounce the  mobility edge. ENAQT magnitude, defined by $\mathcal J/\mathcal J_0$, is measured by comparing a current $\mathcal J$ relative to the one without dephasing, $\mathcal J_0\equiv\mathcal J(\gamma=0)$; ENAQT is achieved when $\mathcal J/\mathcal J_0>1$. This quantity is chosen to give a fair comparison of the current enhancement over ($\alpha$,$\lambda$) variation since for large $\alpha$ or $\lambda$, in which localized eigenstates may exist, the undephased current is already relatively tiny.

\begin{figure}
	\includegraphics[width=\columnwidth]{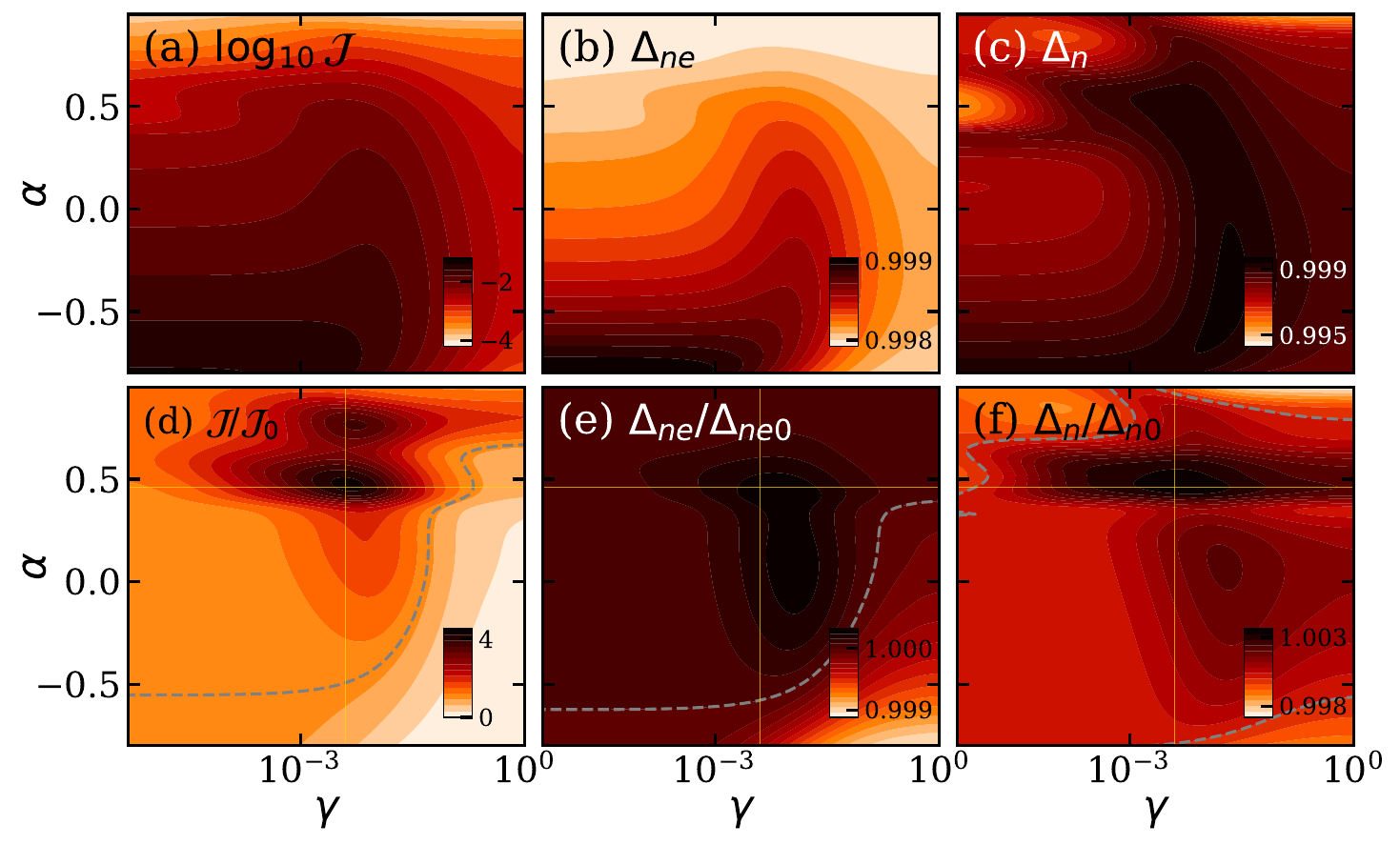}
	\caption{\label{fig:comp}Comparison between (a) current $\mathcal J$ (in logarithmic scale), (b) population spread w.r.t. the exit site, $\Delta_{n\text e}$, (c) inverted population variance, $\Delta_n$, and their relative values (d)--(f) which is normalized by their undephased value ($\gamma=0$). Horizontal and vertical lines indicates the position of $\max(\mathcal J/\mathcal J_0)$, and dashed lines separate the region where the respective relative quantity, (d)--(f), is above $1$. Here $\lambda=0.5$, $\phi=\pi/3$, $T_h=10^{3}$, $T_c=10^{-1}$, and $T_\gamma=10^{-2}$.}
\end{figure}

Before discussing about the uniformization mechanism, which gives rise to ENAQT, we need to assess a couple of population spreading measures exist in the literature, $\Delta_{n\text e}=1-\big(\frac{1}{N}\sum_i n_i - n_\text{ext}\big)^2$ \cite{zerah2018universal} and $\Delta_n = 1 - \frac{1}{N}\sum_i (n_i - \bar n)^2$ \cite{[{We invert the population variance from }] zerah2020effects}, where $n_\text{ext}=\sum_k |S_{Nk}|^2 p_k$ is the (site basis) exit site population, and $p_k=\langle \eta^\dag_k \eta_k \rangle$ is the eigenbasis population. Figure \ref{fig:comp} compares the current $\mathcal{J}$ and its relative value $\mathcal{J}/\mathcal{J}_0$ to both measures, along with their relative values. Apparently, in our case the spread with respect to the exit site, $\Delta_{n\text e}$, correlates better with $\mathcal J$ than the population variance $\Delta_n$, albeit for $\alpha>0.8$ it does not fully sensitive to the change in current since its relative spread, $\Delta_{n\text e}/\Delta_{n\text e0}$ (subscript $0$ means the value for $\gamma=0$), does not properly contain the secondary $\mathcal J/\mathcal J_0$ peak in the high $\alpha$ regime. For other value of $\lambda$ that contains only one ENAQT peak, $\Delta_{n\text e}/\Delta_{n\text e0}$ fits well to estimate the peak location while $\Delta_n/\Delta_{n0}$ does not.

\begin{figure}
	\includegraphics[width=\columnwidth]{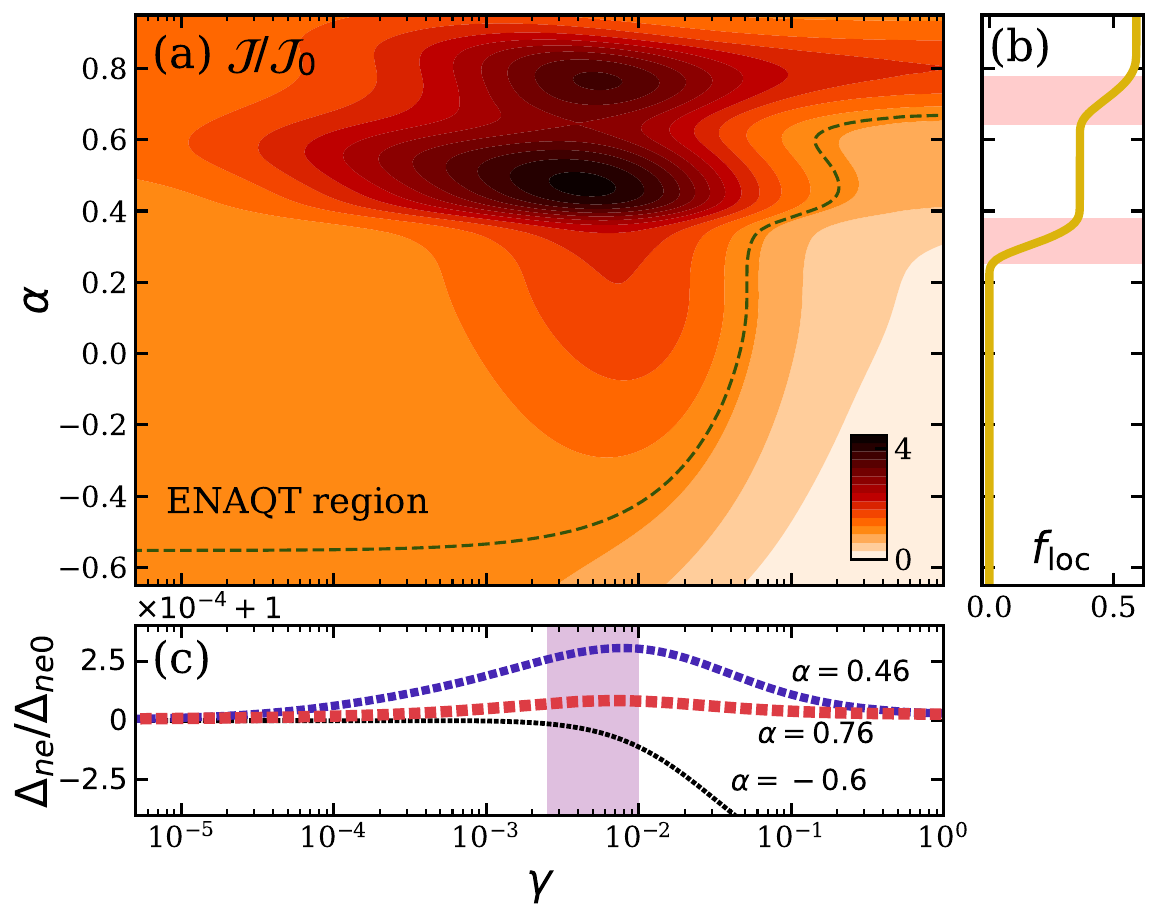}
	\caption{\label{fig:J_norm}(a) Contour plot of relative current $\mathcal J/\mathcal J_0$ as a function of $\alpha$ and $\gamma$. The dashed line corresponds to $\mathcal J/\mathcal J_0=1$ and thus separates the regions with and without ENAQT. (b) Fraction of localized eigenstates $f_\mathrm{loc}$ for each $\alpha$ values. The shaded areas highlight the transition points at which $f_\mathrm{loc}$ varies significantly. (c) Relative spreading, $\Delta_{n\text e}/\Delta_{n\text e0}$, as a function of $\gamma$ for several important $\alpha$'s. The shaded area highlights the location where ENAQT is maximized. The parameters are the same as in Fig. \ref{fig:comp}.}
\end{figure}

The parameter region in which ENAQT exists is illustrated more clearly in Fig. \ref{fig:J_norm}(a). For $\alpha=-1$ (uniform chain), $\mathcal J/\mathcal J_0$ is always decreasing monotonically. This behavior is expected because for an end-to-end transport, ENAQT does not exist unless static disorder is introduced. Meanwhile, for $-1<\alpha<0$ ENAQT exists although minuscule in magnitude. The emergence of global peak at $\alpha=0.46$ (and secondary peak at $\alpha=0.76$) is connected to the existence of localized eigenstates due to mobility edge, which we will elaborate below. For larger disorder strength $\lambda$, the number of localized eigenstates grows for $\alpha\geq0$ and the secondary peak becomes the primary. Eventually, there will be only one peak when $\lambda$ is nearer to $1$. In thermodynamic limit, $N\rightarrow\infty$, the ENAQT peak will be shifted to $\gamma\rightarrow0$, meaning that the current vanishes due to dephasing. 

To understand the role of the mobilty edge, we study the relative current together with localization properties of the GAAH model. This is simply revealed by the fraction of localized eigenstates $f_\mathrm{loc}$. A state is localized if its energy is, for $\alpha>(<)\;0$, larger (lower) than the SPME energy $E_\mathrm{ME}$. By comparing Fig. \ref{fig:J_norm}(a) and (b), we observe that the contour width of ENAQT region varies quite significantly at the transition points of $f_\mathrm{loc}$, that is highlighted by the shaded area just below $\alpha=0.4$ and $\alpha=0.8$. These transitions give rise to the two ENAQT peaks located near the respective $\alpha$ cross sections.

Next, we observe the cooperation between population uniformization and the fraction of localized eigenstates which determines the fate of ENAQT. In Fig. \ref{fig:J_norm}(c) we use $\Delta_{n\text e}/\Delta_{n\text e0}$ to measure the relative spreading. At $\alpha=-0.6$ (outside ENAQT region), it is monotonically decreasing, at $\alpha=0.46$ (global peak) and at $\alpha=0.76$ (secondary peak) it peaks at $\gamma\sim 10^{-2}$ although at the latter $\Delta_{n\mathrm e}$ is much less sensitive as at the former. These peaks correlates with the two $f_\text{loc}$ transitions. In $f_\text{loc}$ plateaux, as can be infered from Fig. \ref{fig:J_norm}(c), the maximum of relative spread keeps decreasing after the first $f_\text{loc}$ transition. Thus the fate of ENAQT there is dominantly controlled by the localized eigenstates. If a NESS contains dominant eigenstates that is effectively suppressed by the dephasing---via the population uniformization mechanism---it will likely generate strong ENAQT although the peak of its population spreading is not prominent. In this sense, $\mathcal{J}/\mathcal{J}_0$ measures \textit{how much the dephasing has uniformized the population} (in site basis), as well as how much it has suppressed localized eigenstates (in eigenbasis).

\begin{figure}
	\includegraphics[width=\columnwidth]{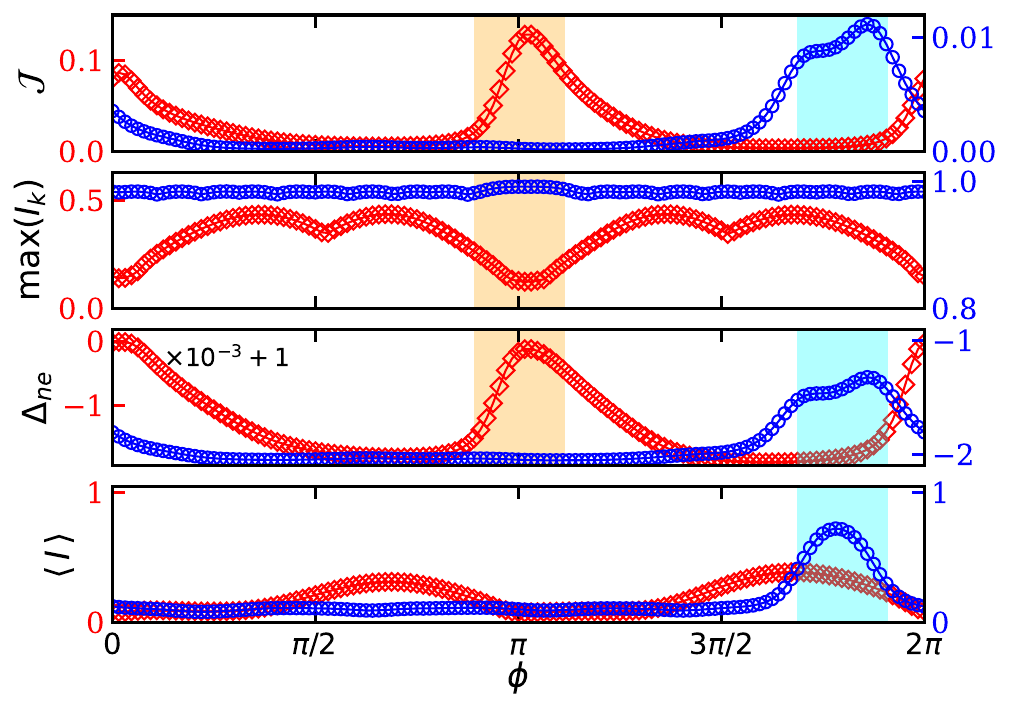}
	\caption{\label{fig:phi}For $\alpha=0.1,\gamma=6\times10^{-3}$ (red $\Diamond$; left y axis) and $\alpha=0.9,\gamma=1$ (blue $\circ$; right y axis): (a) current $\mathcal{J}$ versus phase $\phi$, (b) maximum of coupling between two eigenmodes in presence of dephasing $C_{ik}$, (c) spread $\eta$ as a function of $\phi$, and (d) average inverse participation ratio $\langle I\rangle$. Other parameters are the same as in Fig. \ref{fig:comp}. Shaded areas highlight the correlations between the quantities (see text).} 
\end{figure}

The feature of ENAQT in this model can be explored further by varying the phase $\phi$. This can be achieved in experiments typically by shifting the phase of bichromatic laser field. The phase does not vary the number of localized states, $N f_\mathrm{loc}$, but it alters the spectrum $S_{nk}$. Consequently, the position of the localized states and and the coupling between two modes $C_{ik}$ in Eq. (\ref{coefs}) are changed, resulting a substantial change in ENAQT and localization strength. To study the sensitivity of localization to the dephasing we calculate the average inverse participation ratio (IPR) $\langle I \rangle=\sum I_k p_k$ where $I_k=\sum_n |S_{nk}|^4$ is the IPR for an individual eigenstate.

Figure \ref{fig:phi}(a) shows $\mathcal{J}$ as a function of $\phi$ for two contrasting cases: $\alpha=0.1,\gamma=6\times 10^{-3}$ as the example for maximum ENAQT in the delocalized regime (red $\Diamond$; $f_\mathrm{loc}=0$), and $\alpha=0.9,\gamma=1$ (blue $\circ$; $f_\mathrm{loc}\approx0.6$) for the localized regime. Note that here we do not normalize  $\mathcal{J}$ or $\Delta_{n\text e}$ as they are plotted for specific ($\alpha$,$\gamma$). Both currents peaks at different $\phi$'s but are suppressed in overlapping ranges, i.e., around $\phi=\pi/2$ and $\phi=3\pi/2$. For the delocalized regime, the peak positions can be explained solely by the maximum $I_k$. Turning our attention into the red $\Diamond$ line in Fig. \ref{fig:phi}(b), we see that the $\mathcal J$ is maximized whenever $\max(I_k)$ is small. Proceeding to the blue $\circ$ line, here the maximum eigenstate IPR in Fig. \ref{fig:phi}(b) is no longer sensitive to $\phi$, since there is always at least a localized state with IPR near unity for this case. The corresponding population spread, Fig. \ref{fig:phi}(c), correlates well with both currents, while the average IPR in Fig. \ref{fig:phi}(d) determines the $\mathcal J$ peak only for the localized regime.

\begin{figure}
	\includegraphics[width=0.93\columnwidth]{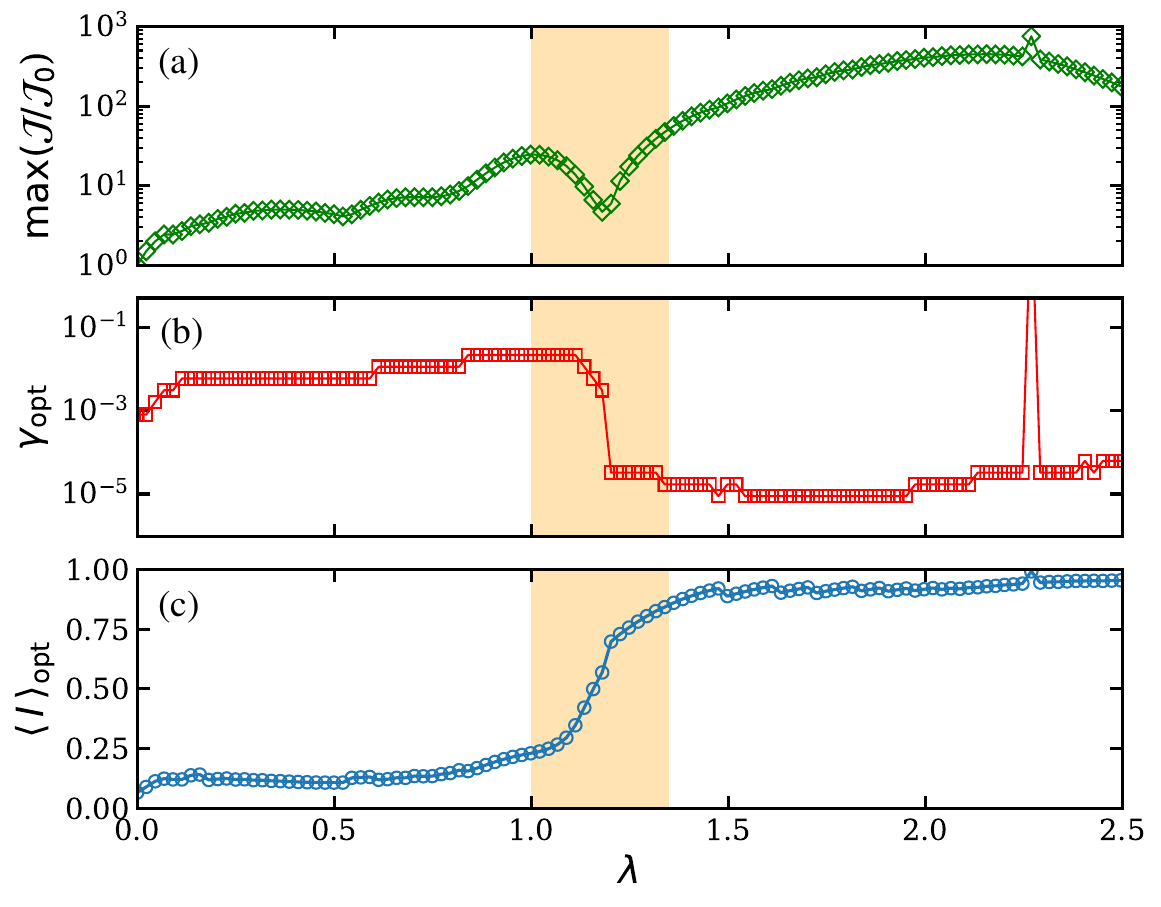}
	\caption{\label{fig:lambda}(a) Maximum ENAQT (occurs at specific $\alpha_\mathrm{opt}$ and $\gamma_\mathrm{opt}$) versus $\lambda$. (b) Dephasing rates at which maximum ENAQT occurs. (c) Average IPR for $\alpha_\mathrm{opt}$ and $\gamma_\mathrm{opt}$. Highlighted area indicates the values at localization transition. Other parameters are the same as in Fig. \ref{fig:comp}.}
\end{figure}

We continue the discussion by including the effect of localization transition. In Fig. \ref{fig:lambda}(a) we plot the maximum ENAQT (achieved at certain $\alpha_\text{opt}$ and $\gamma_\text{opt}$) with $\lambda$. It is possible to obtain $\mathcal{J}/\mathcal{J}_0\approx 400$ at $\lambda\approx 2.1$ for $\gamma_\text{opt}\approx 10^{-5}$. However, further increase in disorder strength reduces the maximum ENAQT as all eigenstates are maximally localized as indicated by $\langle I\rangle_\mathrm{opt}\equiv\langle I\rangle(\alpha_\mathrm{opt},\gamma_\mathrm{opt})$ near unity in Fig. \ref{fig:lambda}(c) and the fact that all eigenstates are localized for $\lambda\gg 1$ and positive $\alpha$. The localization transition from $\lambda=1$ to $\lambda=1.5$ is followed by a steep increase in the system sensitivity to dephasing by orders of magnitude, whereas giving a sharp transition for the maximum ENAQT. In addition, small increments in  Fig. \ref{fig:lambda}(b) for $\lambda=0$ to $\lambda=1$ are followed by a series of local maxima in $\max(\mathcal J/\mathcal J_0)$. The  maximum ENAQT along each $\gamma_\text{opt}$ plateau occurs at $\alpha$ near the $f_\mathrm{loc}$ transition. Interestingly, for $\gamma=0$ the current may be enhanced in presence of nonzero disorder, see Appendix \ref{app:2}. This is in agreement with the result from Ref. \cite{zerah2020effects}.

\begin{figure}
	\includegraphics[width=\columnwidth]{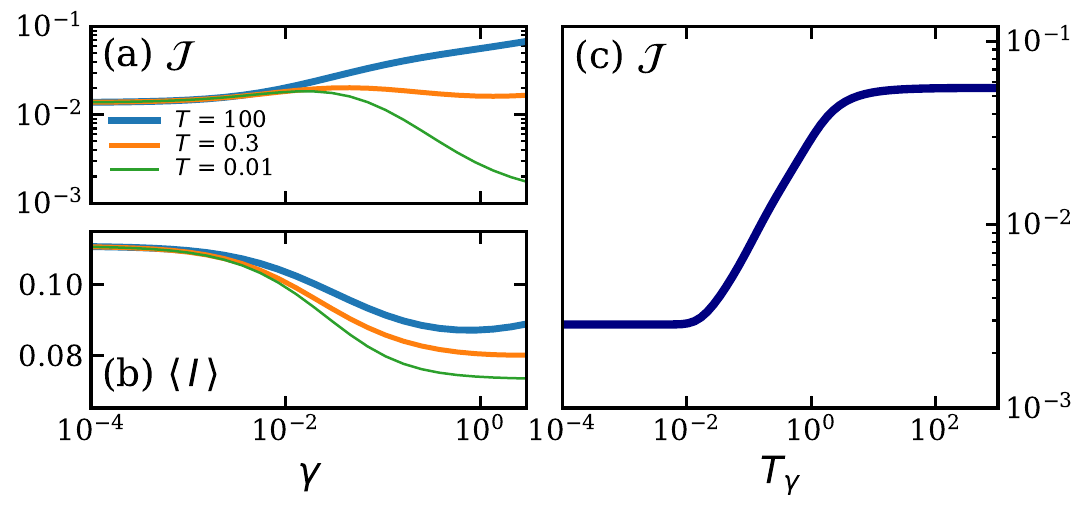}
	\caption{\label{fig:temp}(a) Current and (b) average IPR versus dephasing rate $\gamma$ for various dephasing temperature. (c) Current as a function of dephasing temperature. The parameters are $\lambda=0.4$, $\alpha=0$, $\gamma=0.3$, $T_h=10^3$, and $T_c=10^{-1}$.}
\end{figure}

To complete the phenomenology of this system, we investigate the temperature dependence of the current. Given the step-like behavior of the Fermi-Dirac occupation $n_\gamma(T_\gamma)$, the current will also show a steep transition with respect to dephasing temperature. For $T_\gamma=0$ the matrix $C_{ik}$ in Eq. (\ref{NESS}) is only filled upper diagonally, while for $T_\gamma\rightarrow\infty$ it becomes symmetric. Thus in low temperature the population transfer between eigenmodes occurs only from higher to lower energy, while for high temperature it occurs symmetrically, which in turn affects the high dephasing regime differently, see Appendix \ref{app:3}. We analyze this in Fig. \ref{fig:temp}. Previously we work with $T_\gamma$ below the transition. For $T_\gamma=0.3$ (at transition; see Fig. \ref{fig:temp}(c)) and $T_\gamma=100$, the large dephasing regime becomes strongly enhanced as can be seen in Fig. \ref{fig:temp}(a). This is followed by the increase in average IPR in Fig. \ref{fig:temp}(b). If we use bosonic bath instead, there will be no sharp transition.

\section{CONCLUSIONS}
We have studied how the presence of dephasing noise in a model with SPME coupled to spin baths can manifest strong ENAQT. The peaks of ENAQT correlates with the fraction of localized eigenstates due to mobility edge and the maximum population spreading. Strong ENAQT emerges due to the existence of localized eigenstates that are effectively suppressed by dephasing by virtue of the population uniformization mechanism. The enhancement can be improved further by tuning the phase (the location of localized eigenmodes), disorder strength, and dephasing temperature.

It should be noted that the current is actually higher in the low $\alpha$ regime---with less fraction of localized eigenmodes. However, if disorder or mobilty edge are unavoidably present, or if in an engineered system the mobility edge is desired, the intermediate dephasing regime provides the best performance. We briefly note that if particle interaction is included, ENAQT features should similarly persist since many-body localization occurs in the Hamiltonian \cite{li2015many,modak2015many}.

\begin{acknowledgments}
F.P.Z. thanks LPPM ITB for Research Funding P3MI ITB 2020. The numerical results were obtained using code written in \verb+NumPy+ \cite{harris2020array} and \verb+QuTiP+ \cite{johansson2013qutip}, and the figures were made using \verb+matplotlib+ \cite{hunter2007matplotlib}. 
\end{acknowledgments}

\appendix

\section{DERIVATION OF THE MASTER EQUATION AND ITS STEADY STATE}\label{app:1}
For the microscopic derivation of the master equation, we use the Born-Markov and secular approximation to trace out the bath and write the Lindblad dissipator. This is most readily done using the eigenoperators method \cite{breuer2002theory}. We first write the interaction Hamiltonians in eigenbasis $\eta_k$ where $a_i=\sum_k S_{ik}\eta_k$ ($i=\{l,n\}$) and $S$ is the $N\times N$ unitary matrix which diagonalizes $H_S$,
\begin{eqnarray}\label{Hints}
	H_{SB_1} &=& \sum_{l,\mu,k} g_\mu^{(1)} (S_{lk}\eta_k+S_{kl}^*\eta_k^\dag)(\sigma^+_{\mu,l}+\sigma^-_{\mu,l}), \\
	H_{SB_2} &=& \sum_{n,\mu,k,k'} g_\mu^{(2)} |S_{nk}|^2 \eta_k^\dag \eta_{k'} (\sigma^+_{\mu,n}+\sigma^-_{\mu,n}).
\end{eqnarray}
In terms of the eigenoperators of $H_S$, we can write for each interaction $H_{SB_{(1,2)}}=\sum_{\omega,k} A_{i,k}(\omega) \otimes B_{i}$. Here the decomposed operator $A_{i,k}(\omega)\equiv\sum_{\epsilon'-\epsilon=\omega}\Pi(\epsilon)A_{i,k}\Pi(\epsilon')$ acts on the system, where $\Pi(\epsilon)$ is the projector onto eigenspace belonging to the eigenvalue $\epsilon$ of $H_S$, and  $B_{i}$ acts on the bath. The operator $A_{i,k}(\omega)$ is chosen to satisfy
\begin{equation}
	[H,A_{i,k}(\omega)]=-\omega A_{i,k}(\omega).
\end{equation}
It follows that, for the respective baths, these eigenoperator will be
\begin{eqnarray}
	A_{l,k}(\omega)&=& S_{lk}\eta_k\; \delta_{\omega,+\epsilon_k}+S_{kl}^*\eta_k^\dag\; \delta_{\omega,-\epsilon_k}, \\
	A_{n,k}(\omega)&=& S_{nk} S_{nk'} \eta_k^\dag\eta_{k'}\; \delta_{\omega,\epsilon_{k'}-\epsilon_k}.
\end{eqnarray}
while $B_i = \sum_\mu g_\mu^{(1,2)} (\sigma_{\mu,i}^+ + \sigma_{\mu,i}^-)$ for both $i=\{l,n\}$ [see Eq. (\ref{Hints})]. Intuitively speaking, the coupling $(\eta_k+\eta_k^\dag)$ and $\eta_k^\dag \eta_k$ to the bath induces transitions in the system with the allowed energies $\omega$.

For the each of above eigenoperators, the corresponding Lindblad dissipator is
\begin{eqnarray}
	\mathcal D[A_{i,k}] \,\rho &=& \sum_\omega \Gamma_i (\omega)\Big[A_{i,k}(\omega)\rho A_{i,k}^\dag(\omega) \nonumber \\ && - \frac{1}{2}\{A_{i,k}^\dag(\omega)A_{i,k}(\omega),\rho\}\Big],
\end{eqnarray}
where
\begin{equation}
	\Gamma_i(\omega)=\int_{0}^{\infty}dt\; e^{i\omega t} \mathrm{Tr}_B\left[(e^{iH_St}B_i e^{-iH_St})\, B_i\, \frac{e^{-H_B/T_i}}{\mathrm{Tr}(e^{-H_B/T_i})}\right]
\end{equation}
is the half sided Fourier transform of bath correlation functions. Thus, for $SB_1$ and $SB_2$ we get
\begin{equation}
	\Gamma_i(\omega) = \begin{cases}
		J_i(\omega)[1-n_i(\omega)], &\text{if $\omega>0$}\\
		J_i(-\omega)n_i(-\omega), &\text{if $\omega<0$}
	\end{cases}
\end{equation}
where $n_i(\omega)=(e^{\beta_i\omega}+1)^{-1}$ is the spin occupation number for $SB_1$ ($SB_2$) bath coupled to site $l$ ($n$), and $J_i(\omega)=\sum_\mu \pi|g_\mu^{(1,2)}|^2 \delta(\omega-\epsilon_\mu)$ is the spectral density. for $SB_2$, we denote the spectral density as $\gamma_n(\omega)$. Thus, by writing for $\omega>0$ and $\omega<0$ in a unified way, we can write the full dissipators as
\begin{eqnarray}
	\mathcal D_{SB_1} \rho &=& \sum_{l,k} |S_{lk}|^2 J_1(|\epsilon_k|)\Big[\left(1-n_l(\epsilon_k)\right)(\eta_k\rho\eta_k^\dag  \nonumber \\ && -\frac{1}{2}\{\eta_k^\dag\eta_k,\rho\}) + n_l(\epsilon_k)(\eta_k^\dag\rho\eta_k - \frac{1}{2}\{\eta_k\eta_k^\dag,\rho\})\Big], \\
	\mathcal D_{SB_2} \rho &=& \sum_{n,k,k'} |S_{nk}|^2 |S_{nk'}|^2  \Big[J_2(\epsilon_{k'}-\epsilon_k)(1-n_\gamma(\epsilon_{k'}-\epsilon_k)) \nonumber \\ && +J_2(\epsilon_k-\epsilon_{k'}) n_\gamma(\epsilon_k-\epsilon_{k'})\Big]\nonumber (\eta_k^\dag\eta_{k'}\,\rho\,\eta_{k'}^\dag\eta_{k} \nonumber \\ &&-\frac{1}{2}\left\{\eta_k^\dag\eta_{k'},\rho\right\}, \label{deph}
\end{eqnarray}
where we set constant $J_2(\omega)=\gamma$ for $\omega>0$ and $J_2(\omega)=0$ elsewhere. Note that by $n_\gamma$ we mean the occupation due to index $n$ ($SB_2$ bath).

\begin{figure}
	\centering
	\includegraphics[width=0.8\columnwidth]{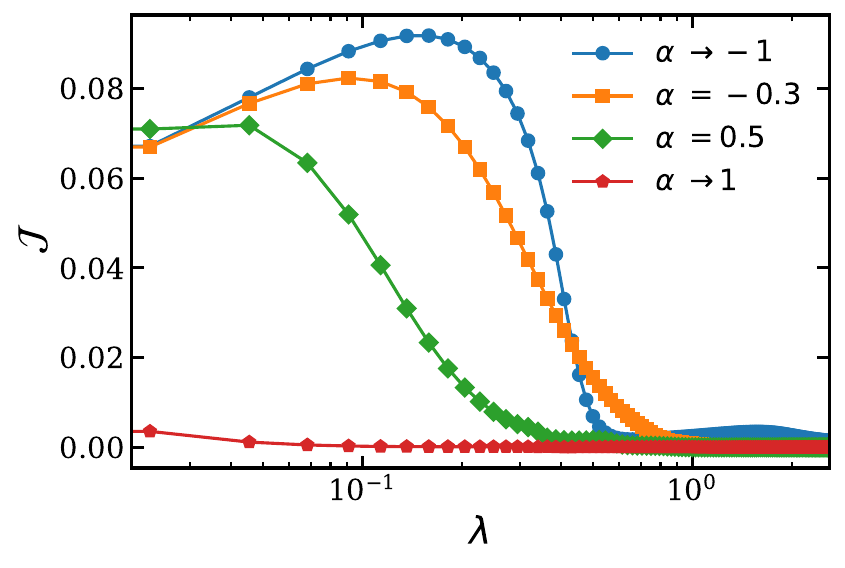}
	\caption{\label{fig:dis}Energy current $\mathcal J$ as a function of disorder strength $\lambda$ for various on-site potential deformation parameter $\alpha$ without the presence of dephasing noise ($\gamma=0$). Other parameters are set the same as in Fig. \ref{fig:comp}.}
\end{figure}

To find the nonequilibrium steady state (NESS), we consider the adjoint master equation,  $\mathcal D_{i,k}[A_{i,k}]^\dag\, O = A_{i,k}^\dag O A_{i,k} - 1/2\{A_{i,k}^\dag A_{i,k},O\}$, by replacing $O=\eta_k^\dag\eta_{k'}$ and solving $\langle dO/dt \rangle = 0$. The solution can be found by using only the fundamental commutation relations, which simplifies the adjoint dissipators considerably,
\begin{eqnarray}
	\mathcal D^\dag_{SB_1} \Big(\eta_k^\dag\eta_{k'}\Big) &=&\sum_l |S_{lk}|^2 \Big(n_l(\epsilon_k)\delta_{kk'} \nonumber \\&& -[1-2n_l(\epsilon_k)]\eta_k^\dag\eta_{k'}\Big), \\
	\mathcal D^\dag_{SB_2} \left(\eta_k^\dag\eta_{k}\right) &=& \eta_k^\dag \eta_{k'} \Bigg[\sum_i(C_{ik}-C_{ki}) \eta^\dag_i\eta_i -\sum_i C_{ik}\Bigg] \nonumber \\&& + \sum_i C_{ik} \eta_i^\dag\eta_i, \label{D2}
\end{eqnarray}
where
\begin{equation}
	C_{ik}=\tilde\gamma_{ik} \sum_n |S_{ni}|^2|S_{nk}|^2,
\end{equation}
and $\tilde\gamma_{ik}=J_2(\omega)\big(1+n_\gamma(\omega)\big)+J_2(-\omega)n_\gamma(-\omega)$ with $\omega=\epsilon_{i}-\epsilon_k$. The first term in the square bracket of Eq. (\ref{D2}) vanishes since we only consider one particle manifold. The NESS is given by Eq. (\ref{coefs}). The reader can verify that the diagonal element of the NESS, $k\neq k'$, is zero. Once the NESS ($\langle\eta^\dag_k\eta_k\rangle$) is found, we can proceed to calculate the energy current $\mathcal J$ with a similar calculation, resulting in Eq. (\ref{J}).

\section{THE GAAH POTENTIAL AND EFFECT OF DISORDER STRENGTH}\label{app:2}
First we need to differentiate the role of $\alpha$ and $\lambda$ to the on-site potential $V_n$. Disorder $\lambda$ varies the potential height by a mere scaling, while the mobility edge parameter $\alpha$ deforms the potential by amplifying the variance between the peaks; this in turn creates different localization properties for different energy levels. For $\alpha\rightarrow1$, the potential approaches singularity in some places. This typically induces errors in numerical calculation and should be avoided.


There is an interesting effect of disorder strength $\lambda$. Nonzero disorder may enhance the transport, even in case of $\gamma=0$. This is in line with the result in Ref. \cite{zerah2020effects}. This implies that the maximum current is achieved at a finite disorder strength. As $\alpha$ increases from $-1$ to $1$, the energy current peak becomes less pronounced, see Fig. \ref{fig:dis}.

\section{TEMPERATURE DEPENDENCE OF THE CURRENT}\label{app:3}
\begin{figure}
	\includegraphics[width=\columnwidth]{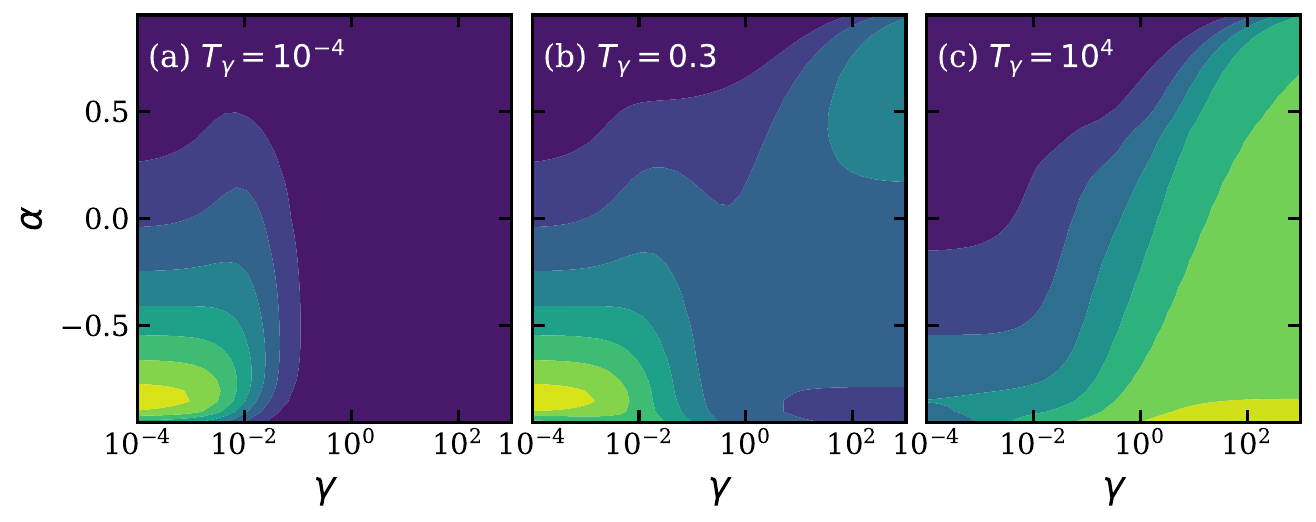}
	\caption{\label{fig:T}Contour plot of $\mathcal J$ (unnormalized) as a function of $\alpha$ and $\gamma$ for (left) low $T_\gamma$, (center) $T_\gamma$ at the transition in Fig. \ref{fig:temp}, and (right) high $T_\gamma$. Greenish contour indicates larger $\mathcal J$.}
\end{figure}

The results this paper are mainly produced by taking low dephasing temperature $T_\gamma$ and we briefly discuss the effect of increasing the temperature in Fig. \ref{fig:temp}. In Fig. \ref{fig:T} we depict the temperature dependence of the $(\alpha,\gamma)$ contour. Recall that there is a sharp transition around $T_\gamma=1$ due to the step-like behavior of $n_\gamma(\omega)$ in Fermi-Dirac statistics of the dephasing bath. As $T_\gamma$ passes the transition regime, a new regime with high $\mathcal J$ opens up in strong $\gamma$'s. At high $T_\gamma$ limit the maximum ENAQT is totally shifted to the high dephasing regime. Note that this should be taken with caution because with the high value of $\gamma$, global master equation may give compromised results.

\bibliography{dwiputra_ENAQT} 

\clearpage

\end{document}